\documentclass[sigconf]{acmart}

\usepackage{multirow}
\usepackage{cleveref}
\usepackage{tikz}
\usepackage{svg}
\usetikzlibrary{shapes.geometric, arrows, shapes.symbols, decorations.pathreplacing, calligraphy}

\tikzstyle{process} = [rectangle, minimum width=4cm, minimum height=2cm, text centered, draw=black]
\tikzstyle{arrow} = [thick, ->, >=stealth]

\usepackage{varwidth}

\AtBeginDocument{%
  \providecommand\BibTeX{{%
    \normalfont B\kern-0.5em{\scshape i\kern-0.25em b}\kern-0.8em\TeX}}}

\setcopyright{acmcopyright}
\copyrightyear{2018}
\acmYear{2018}
\acmDOI{XXXXXXX.XXXXXXX}

\acmConference[Conference acronym 'XX]{Make sure to enter the correct
  conference title from your rights confirmation emai}{June 03--05,
  2018}{Woodstock, NY}
\acmPrice{15.00}
\acmISBN{978-1-4503-XXXX-X/18/06}

\copyrightyear{2022} 
\acmYear{2022} 
\setcopyright{acmlicensed}\acmConference[AIES'22]{Proceedings of the 2022 AAAI/ACM Conference on AI, Ethics, and Society}{August 1--3, 2022}{Oxford, United Kingdom}
\acmBooktitle{Proceedings of the 2022 AAAI/ACM Conference on AI, Ethics, and Society (AIES'22), August 1--3, 2022, Oxford, United Kingdom}
\acmPrice{15.00}
\acmDOI{10.1145/3514094.3534128}
\acmISBN{978-1-4503-9247-1/22/08}

\begin{document}

\title{A Meta-Analysis of the Utility of Explainable Artificial Intelligence in Human-AI Decision-Making}

\author{Max Schemmer}
\authornote{Both authors contributed equally to this research.}
\email{max.schemmer@kit.edu}
\affiliation{%
  \institution{Karlsruhe Institute of Technology}
  \city{Karlsruhe}
  \country{Germany}
}

\author{Patrick Hemmer}\authornotemark[1]
\email{patrick.hemmer@kit.edu}
\affiliation{%
  \institution{Karlsruhe Institute of Technology}
  \city{Karlsruhe}
  \country{Germany}}

\author{Maximilian Nitsche}
\email{maximilian.nitsche@student.kit.edu}
\affiliation{%
  \institution{Karlsruhe Institute of Technology}
  \city{Karlsruhe}
  \country{Germany}
}

\author{Niklas Kühl}
\email{niklas.kuehl@kit.edu}
\affiliation{%
 \institution{Karlsruhe Institute of Technology}
 \city{Karlsruhe}
 \country{Germany}}

\author{Michael Vössing}
\email{michael.voessing@kit.edu}
\affiliation{%
  \institution{Karlsruhe Institute of Technology}
  \city{Karlsruhe}
  \country{Germany}}



\begin{abstract}
Research in artificial intelligence (AI)-assisted decision-making is experiencing tremendous growth with a constantly rising number of studies evaluating the effect of AI with and without techniques from the field of explainable AI (XAI) on human decision-making performance. However, as tasks and experimental setups vary due to different objectives, some studies report improved user decision-making performance through XAI, while others report only negligible effects. Therefore, in this article, we present an initial synthesis of existing research on XAI studies using a statistical meta-analysis to derive implications across existing research. We observe a statistically positive impact of XAI on users' performance. Additionally, the first results indicate that human-AI decision-making tends to yield better task performance on text data. However, we find no effect of explanations on users' performance compared to sole AI predictions. Our initial synthesis gives rise to future research investigating the underlying causes and contributes to further developing algorithms that effectively benefit human decision-makers by providing meaningful explanations. 
\end{abstract}

\ccsdesc[500]{Human-centered computing~Empirical studies in HCI}
\ccsdesc[300]{Computing methodologies~Artificial intelligence}

\keywords{Explainable Artificial Intelligence, Decision-Making, Empirical Studies, Meta-Analysis}



\maketitle


\section{Introduction}
Over the last years, the rapid developments in artificial intelligence (AI) have increased its use in many application domains. In this context, AI's continuously rising capabilities have surpassed human performance in an increasing number of tasks, such as playing poker \cite{Brown2019SuperhumanAF},  go \cite{Silver2018AGR}, or correctly recognizing various categories of interest in images \cite{He2015DelvingDI}. Due to these remarkable developments, AI is increasingly applied to support decision-makers in an increasing number of domains, such as medicine \cite{McKinney2020InternationalEO,Wu2020DeepNN}, finance \cite{day2018ai}, law \cite{Kleinberg2018HumanDA} or manufacturing \citep{stauder2021ai}.  

To offer decision-makers meaningful support, AI models are expected to provide accurate predictions and a notion of how a particular decision has been derived. In particular, explaining the rationale behind an algorithmic decision should enable domain experts to learn when to trust the recommendations of the AI and when to question it \cite{zhang_efect_2020}. This requirement fueled the continuous development of explainability techniques from the field of explainable AI (XAI), intending to make the decision-making process of black-box AI models more transparent and, thus, comprehensible for domain experts \cite{adadi_peeking_2018}. Common approaches include among others feature importance-based \cite{ribeiro_why_2016}, example-based \cite{Cai2019TheEO}, or rule-based methods \cite{ribeiro2018anchors}. A better understanding of how the AI's decision was derived should subsequently enable the user to appropriately rely on the AI's suggestions on a case-by-case basis \cite{bansal_does_2020, Lee2004TrustIA}. For instance, explanations contradicting the AI's prediction could signify the user to become skeptical, consequently considering the AI prediction less in the final decision-making process. 

With the ongoing development of XAI techniques, researchers have started to evaluate AI with and without explanations to assess whether their utility for better decision-making can be quantified \citep{bansal_does_2020,bucinca_proxy_2020,bucinca_trust_2021,carton_feature-based_2020,chu_are_2020,green2019principles,hase_evaluating_2020,hemmer2022effect,lai_human_2019,lai_why_2020,liu_understanding_2021,yeung_sequential_2020,zhang_efect_2020,van_der_waa_evaluating_2021}. Whereas some researchers identify a benefit of XAI-based decision support in user studies \cite{bucinca_proxy_2020,lai_why_2020}, others find only negligible evidence \citep{carton_feature-based_2020,liu_understanding_2021}, with the underlying causes remaining partly unexplored \citep{hemmer2021human,schoeffer2022relationship}. Therefore, in this article, we aim to clarify the current ``snapshot'' of the utility of XAI-based decision support. We conduct a meta-analysis of user studies identified in a structured literature review to shed light on the effect of XAI-assisted decision-making on user performance. In detail, our analysis encompasses studies that allow a comparison between human, AI-, and XAI-assisted task performance. Our initial findings are the following: First, on average, XAI-assisted decision-making enhances human task performance compared to no assistance at all. However, we find no additional effect of explanations on users' performance in XAI-assisted decision-making compared to isolated AI predictions, which raises questions on how to further develop current XAI methods that improve users' task performance. Second, we find that distinct data types affect user performance differently. In this context, human-AI decision-making turns out to be more effective on text data compared to tabular data.

The remainder of this article is structured as follows. In \Cref{sec:RW}, we first outline related work in the context of human-AI decision-making. In \Cref{sec:methodology}, we describe the methodological approach of our meta-study. Subsequently, we present the results of the meta-study, including a subgroup analysis in \Cref{sec:results}. In this context, we provide additional qualitative insights on an individual level. We outline the current limitations of our work in \Cref{sec:limitations}, followed by a discussion on relevant implications that result from these findings for the future development of XAI algorithms in \Cref{sec:discussion}. Finally, \Cref{sec:conclusion} concludes our work.

\section{Related Work}\label{sec:RW}

Over the last years, research has focused on developing algorithms that provide explanations for AI predictions \cite{adadi_peeking_2018,das2020opportunities}. By now, these algorithms are increasingly employed in a growing number of practical use cases such as in manufacturing \cite{Senoner2021UsingEA,treiss2020uncertainty}, medicine \cite{Pennisi2021AnEA}, or the hospitality industry \cite{voessing2022}. Usually, XAI is utilized in scenarios that involve humans-in-the-loop processes. The underlying idea is that humans will benefit from the AI's suggestion if it is accompanied with an explanation. Therefore, a constantly rising number of studies has started to analyze the effects of explanations in behavioral experiments \cite{hemmer2021human}. In these experiments, many different target variables are taken into consideration, e.g., whether humans are capable of predicting what a model would recommend (proxy tasks) \cite{bucinca_proxy_2020,chandrasekaran_explanations_2018,hase_evaluating_2020} or whether explanations support them in model debugging \cite{adebayo2020debugging,Kaur2020InterpretingIU}.

In the scope of this study, we explicitly focus on AI-assisted decision-making---a setting in which an AI supports a human with the goal of improving the decision-making quality. The prediction of the AI might be accompanied by additional information, e.g., about its prediction uncertainty or different types of explanations. After receiving the AI's advice, the human decision-maker is responsible for making the final decision. A scenario, which is often also required from a legal perspective, as the human needs to make the final decision \cite{bauer2021expl}. By providing either additional information on the AI's prediction uncertainty \cite{nguyen2021effectiveness,zhang_efect_2020} or explanations on how a decision was derived \cite{bansal_does_2020,lai_human_2019}, humans shall be enabled to better question the AI's decision. To develop a deeper understanding of this assumption, research has evaluated the effect of explanations on users' trust and how reliance on AI decisions can be appropriately calibrated \cite{bucinca_trust_2021,kunkel_let_2019,Yu2019DoIT,zhang_efect_2020,bansal_does_2020,schemmer2022influence}. In this context, providing humans not only with the AI's prediction and respective explanations but also with a notion about its global performance can influence the overall team performance \cite{lai_human_2019}. Additional benefits can be found when humans are provided with model-driven tutorials about AI functionality and the task itself \cite{lai_why_2020}. Further work in this line of research has investigated the influence of AI advice in the out-of-distribution setting---instances differing from the distribution used for AI training---on the final human decision \cite{liu_understanding_2021}.   

Besides these factors, the explanation type of an AI prediction can play a decisive role. In this context, research has developed various explainability techniques \cite{adadi_peeking_2018} ranging from feature importance methods \cite{ribeiro_why_2016} over example-based approaches \cite{Cai2019TheEO} to rule-based explanations \cite{ribeiro2018anchors} that have been evaluated in user studies accordingly \cite{hemmer2021human}. However, the current picture emerging from the results of different studies regarding the effects of XAI methods on AI-assisted decision-making performance is not unambiguous. Whereas, e.g., \citet{carton_feature-based_2020} conclude that feature-based explanations do not help users in classification tasks, \citet{hase_evaluating_2020} find some of them to be effective in model simulatability, which refers to the ability to predict the model behavior given an input and an explanation. In this context, further studies demonstrate the utility of explanations \cite{bucinca_proxy_2020}, whereas others find that they can convince humans to follow incorrect suggestions more easily \citep{bansal_does_2020,van_der_waa_evaluating_2021}. 

Of course, ambiguous findings can also be attributed to the specific setups of each study and the different goals pursued by the researchers. We aim to shed light on this ambiguity by conducting a meta-analysis of human-AI decision-making---particularly on the influence of explainability.

\section{Methodology}\label{sec:methodology}

We elaborate on our data collection approach to identify relevant articles, followed by the statistical analysis conducted on the final set of user studies.

\begin{table*}[t]
\centering
\resizebox{\linewidth}{!}{%
\begin{tabular}{llllc}
\hline 
\multicolumn{1}{c}{Source} & \multicolumn{1}{c}{Dataset} & \multicolumn{1}{c}{Datatype}     & \multicolumn{1}{c}{Studies} 
\\  \hline 
     \multirow{2}{*}{\citet{alufaisan_does_2020}}  & 
     COMPAS \cite{propublica}
     & \multirow{2}{*}{Tabular} 
     & AI-, XAI-assisted (Anchor)           
            \\  
     & Census \cite{Dua2019}
     & 
     & AI-, XAI-assisted (Anchor)
      \\ \hline 
     
     \multirow{3}{*}{\citet{bansal_does_2020}}  & 
     LSAT \cite{team2017lsat}
     &     
     \multirow{3}{*}{Text}                     
     & XAI-assisted (Confidence, Explain Top-1 Expert, Explain Top-2 Expert, Adaptive Expert)          
       \\ 
     
     & Book reviews \cite{he2016ups} &
     & XAI-assisted (Confidence, Explain Top-1 AI, Explain Top-2 AI, Adaptive AI, Adaptive Expert)
      \\ 
     & Beer reviews \cite{mcauley2012learning} 
     &
     & XAI-assisted (Confidence, Explain Top-1 AI, Explain Top-2 AI, Adaptive AI, Adaptive Expert)
      \\ \hline 
     
     \citet{bucinca_proxy_2020}  & Fat content prediction \cite{bucinca_proxy_2020}       
     &                          
     Image &     AI-, XAI-assisted (Inductive \& Deductive Explanation)      
                        \\\hline 
     
     \citet{carton_feature-based_2020}  
     & Online toxicity \cite{wulczyn2017ex}        
     & Text 
     & AI-, XAI-assisted (Keyword, Partial, Full Explanation)           
                         \\\hline 
      
            \citet{fugener2021will} &         
      Dog breed ImageNet \cite{russakovsky2015imagenet}&                          
      Image &    AI-, XAI-assisted (Certainty)       
      \\ \hline 

      \citet{lai_why_2020} 
      & Deception detection \cite{ott2013negative,ott2011finding}       
      & Text 
      & XAI-assisted (Signed \& Predicted Label, Signed \& Predicted Label \& Guidelines, Signed \& Predicted Label \& Guidelines \& Accuracy)       
                        \\\hline

      \multirow{3}{*}{\citet{liu_understanding_2021}} 
      & COMPAS \cite{propublica}
      & \multirow{2}{*}{Tabular} 
      & XAI-assisted (Static/Static, Interactive/Static, Interactive/Interactive)          
      \\
      
      & ICPSR \cite{ICPSR} &  & XAI-assisted (Static/Static, Interactive/Static, Interactive/Interactive) \\ 
      & BIOS \cite{bios} & Text & XAI-assisted (Static/Static, Interactive/Static, Interactive/Interactive)  \\ \hline

      \citet{van_der_waa_evaluating_2021} &         
      Diabetes mellitus type 1 \cite{van_der_waa_evaluating_2021} &                          
      Tabular &  AI-, XAI-assisted (Rule-based, Example-based)         
      \\\hline

      \citet{zhang_efect_2020} &         
      Census \cite{Dua2019} &                       
      Tabular &  AI-, XAI-assisted (Confidence, Feature Importance)        
      \\ 
      \hline 
\end{tabular}
}
\caption{Overview of articles that were identified in the structured literature review and are analyzed in this work. All articles are peer-reviewed at the time of the meta-study.}
\label{tab:article_overview}
\end{table*}

\subsection{Data Collection}
For the collection of empirical user studies in the field of XAI, we conducted a structured literature review based on the methodology outlined by \citet{brocke_reconstructing_2009}. In detail, we developed a search string focusing on XAI and behavioral experiments. For both topics, several synonyms were included after an explorative search. Subsequently, the search string was iteratively refined, resulting in the following final search string:  
 
\begin{center}
\textit{
TITLE-ABS-KEY("explainable artificial intelligence" OR XAI OR "explainable AI" OR ( ( interpretability OR explanation ) AND ( "artificial intelligence" OR AI OR "machine learning" ) ) ) AND ( "human performance" OR "human accuracy" OR "user study" OR "empirical study" OR "online experiment" OR "human experiment" OR "behavioral experiment" OR "human evaluation" OR "user evaluation")}
\end{center}

To ensure comprehensive coverage of relevant articles, we chose the SCOPUS database for our initial search \cite{schotten2017brief}. We filtered identified articles according to the following three criteria: an article identified with the search string was included if it (a) conducted at least one empirical user study and (b) reported the task performance as a performance measure for humans and AI- or XAI-assisted decision-making on the same task.

Additionally, we conducted a forward and backward search starting from the articles that fulfill our inclusion criteria. We extracted all individual treatments and outcomes for each article. For instance, if an experiment compared AI- with XAI-assisted decision-making in a between-subject design in two separate treatments, each of them was registered as a separate record in our database. If an article includes multiple experiments, we perform the data extraction process for each experiment separately. We contacted authors by email in case of missing or not reported information in the articles regarding the conducted user studies.

The collected studies vary considerably in terms of tasks, problem settings, and reported performance metrics. Accordingly, we filter our set of studies in the following way: First, we focus on studies assessing classification tasks as they account for the largest subset across all entries in our database. Second, we restrict the subset of relevant studies to those reporting the mean accuracy as the performance measurement in each study since we require a common metric across multiple studies. This ensures that we base our meta-analysis on comparable and interpretable effect sizes. Third, we only include studies that have been conducted as a between-subject design. By excluding studies conducted in a within-subject design, we avoid taking into account the learning effect of participants between treatments that might distort the effect sizes of our analysis. 
\begin{figure}[h]
\centering
\begin{tikzpicture}[node distance=3cm]

\node (process1) [process] {Total records identified (\(n=393\)).};
\node (process2) [process, below of=process1] {\begin{varwidth}{15em} Number of records after filtering according to eligibility criteria and forward/backward search (\(n=33\)).\end{varwidth}};
\node (process3) [process, below of=process2] {\begin{varwidth}{15em} Records after excluding studies with incomplete information and filtering for classification tasks (metric: accuracy) conducted in a between-subject design (\(n=9\)).\end{varwidth}};

\draw[arrow] (process1) -- (process2);
\draw[arrow] (process2) -- (process3);
\end{tikzpicture}
\caption{Flowchart describing the data collection and article selection procedure.}
\label{fig:flowchart}
\end{figure}
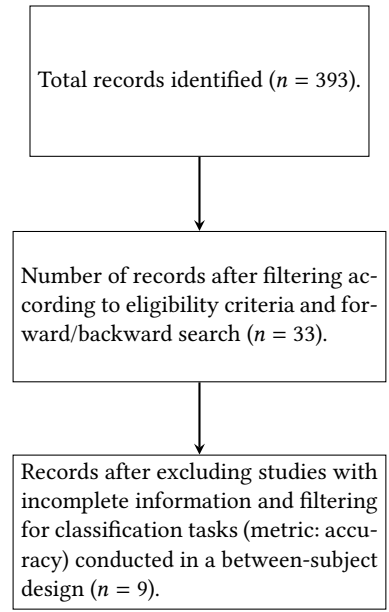

Subsequently, we extract all necessary performance metrics from the articles. Next, we define the case in which the human performs the task without any AI support as human performance. If the human is additionally equipped with AI advice, but without explanations, we call the performance AI-assisted performance. AI assistance with explanations is called XAI assistance, and the resulting performance measure is denoted as XAI-assisted performance. Based on these definitions, we excluded all studies that do not report human performance and either AI-assisted or XAI-assisted performance. Based on the resulting sample, we conduct the following statistical analysis.

\subsection{Statistical Analysis}

For each study, we calculate the effect size as the between-group standardized mean difference (SMD) of the task performance. Furthermore, we report Hedges' \(g\) \cite{hedge} to correct the SMD for a possible upward bias of the effect size when the sample size of a study is small (\(n \leq 20\)). Thus, Hedges' \(g\) is smaller for n$\leq$20 than the uncorrected SMD but approximately the same for larger sample sizes. We obtain the standard deviations from standard errors and confidence intervals for group means reported for each treatment following the procedure outlined in \citet{higgins_cochrane_2019}. 
In case an article encompasses multiple studies with a single control group, we divide the size of this control group by the number of studies to avoid multiple comparisons against the same group \cite{higgins_cochrane_2019}.

For our meta-analytic model and the pooling of effect sizes, we estimate a random-effects model as the setups and populations are considerably heterogeneous between studies. Hence, we calculate the distribution mean of effect sizes instead of estimating and assuming one single true effect size underlying the studies (fixed-effect model \cite{borensteinBasicIntroductionFixedeffect2010}). To assess the between-study heterogeneity variance ${\tau^{2}}$ and its confidence intervals we use the DerSimonian-Laird estimator \cite{dersimonian_meta-analysis_1986} and Jackson's method \cite{jackson_confidence_2016}, respectively. 

Additionally, we conduct a subgroup analysis to provide further insights across current XAI studies. Several studies have discussed that task choice has a strong influence on the experimental outcome \cite{fugener2021will,lai_towards_2021}. This article focuses on the influence of the task's data type. In this context, many researchers have argued about the importance of data types in human-AI decision-making. For example, \citet{fugener2021will} reason that image recognition, in general, is well suited for human-AI decision-making since it is an intuitive task for humans.

\section{Results}\label{sec:results}
We start by presenting the final set of included studies, then outline the meta-study results, including the respective subgroup analyses. Finally, we provide additional qualitative insights on an article level.

\subsection{Data Collection}
As of February 2022, we identified a total number of 393 articles. After applying our inclusion criteria and conducting a forward and backward search, the number of relevant articles is reduced to 33. 

As classification tasks form the largest subset, we focus on this particular prediction problem. After filtering for accuracy as a common metric and removing articles with missing information, e.g., sample size or dispersion measures, we include 9 articles in the meta-analysis and the respective subgroup analyses. \Cref{fig:flowchart} visualizes the entire filtering process. Moreover, \Cref{tab:article_overview} provides an overview of all included articles together with information about each dataset, datatype, and treatments extracted from the articles. Each article contains at least one behavioral experiment conducted with a particular dataset. Each experiment consists of several experimental treatments. The treatment in which humans conducted a task on their own without AI assistance is referred to as a control group. In the following, we denote each treatment as an individual study. Overall, we thereby have a sample size of 44 studies.

\subsection{AI Assistance vs. XAI Assistance} \label{sec:ai-vs-xai} 

\begin{figure*}[h!]
\centering
\includegraphics[width=0.95\linewidth]{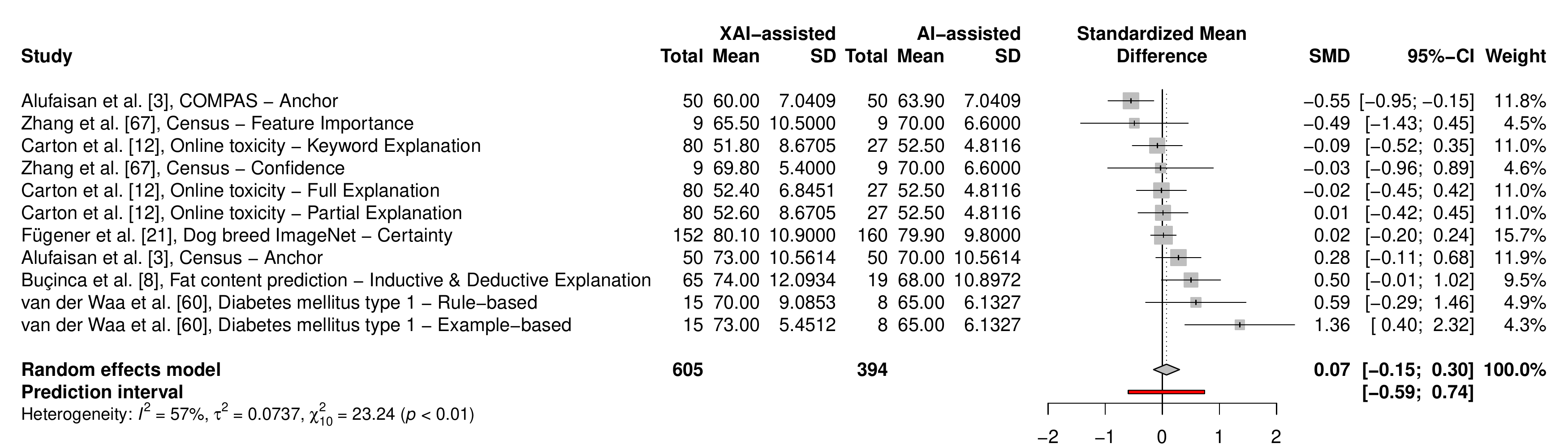}
\caption{Forest plot of the standardized mean difference between AI-assisted and XAI-assisted performance.}
\label{fig:AI-XAI}
\end{figure*}

We start our meta-analysis by investigating AI- and XAI-assisted performance. For this reason, we first focus on all studies that report AI- and XAI-assisted performance, which leads us to a sample of 11 studies and a total number of 999 observations.  

\Cref{fig:AI-XAI} displays the forest plot of the standardized mean difference (SMD) between AI-assisted and XAI-assisted performance. The results of the analysis reveal that, on average, the SMD of all studies that reported AI- and XAI-assisted performance is 0.07 with a 95\% confidence interval (CI) [-0.15, 0.30]. A z-test against the null-hypothesis that the effect size is 0 cannot be rejected (\(z= 0.63, p = 0.53\)). This means we do not find a significant difference between AI-assisted and XAI-assisted performance in our current sample of studies. Regarding heterogeneity, we find an \(I^2\) of 57.00\% (95\% CI [15.70\%, 78.00\%]), which can be considered moderate \cite{higgins_cochrane_2019}. The \(\tau^2\) is 0.07 (95\% CI [0.01, 0.54]) and Q is significantly different from 0 (\(Q=23.24, df=10, p < 0.01\)). Thus, we can reject the null hypothesis that the true effect size is identical in all studies. To provide an intuitive understanding of the heterogeneity, we also report the prediction interval that represents the expected range of true effects in other studies \cite{inthout2016plea}. The prediction interval ranges from -0.59 to 0.74. That means we can expect negative as well as positive effects of XAI assistance in comparison to AI assistance. In summary, on average, XAI-assisted decision-making does not significantly influence the performance of human-AI decision-making in our sample. The highest improvement was measured by \citet{van_der_waa_evaluating_2021}. Contrary, the highest negative impact of XAI is measured by \citet{alufaisan_does_2020}.

\subsection{Human vs. XAI Assistance}\label{sec:xai-vs-human} 
\begin{figure*}[h]
\centering
\includegraphics[width=0.95\linewidth]{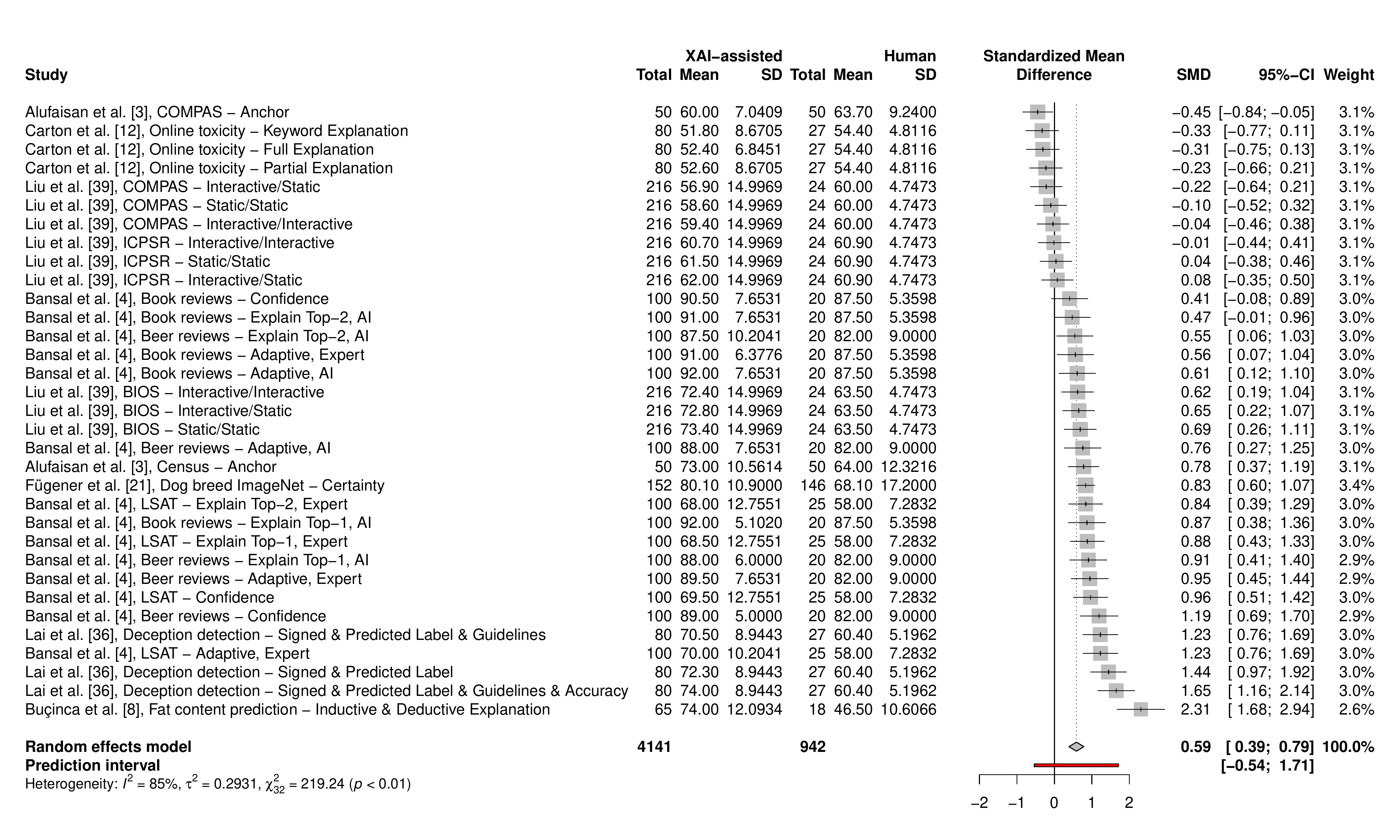}
\caption{Forest plot of the standardized mean difference between human and XAI-assisted performance.}
\label{fig:Human-(XAI)}
\end{figure*}

We analyze the overall effect of XAI in comparison with human performance. Therefore, we filter all studies that report human and XAI-assisted performance. This results in a sample of 33 studies and a total number of 5,083 participants. Based on this sample, we analyze whether XAI-assisted decision-making improves performance compared to humans conducting a task alone. 

\Cref{fig:Human-(XAI)} visualizes the forest plot of the standardized mean difference between human and XAI-assisted performance. The meta-analysis indicates that, on average, XAI assistance increases task performance by 0.59 SMD as compared to humans conducting the tasks alone. The 95\% CI of the SMD ranges from 0.39 to 0.79. As this range does not include an effect size of 0 and a z-test is significant (\(z = 5.73, p < 0.0001\)), we can reject the null hypothesis concluding that, on average, XAI-assisted decision-making improves human task performance.   
Looking at heterogeneity, we can reject the null hypothesis that the true effect size is identical in all studies (\(Q = 219.24, df = 32, p < 0.0001\)). Moreover, \(I^{2}\) is 85.40\% with a 95\% CI of 80.50\% to 89.10\%. The estimated \(\tau^{2}\) is 0.29 (95\% CI [0.18, 0.59]). Thus, the level of heterogeneity can be considered substantial \cite{higgins_cochrane_2019}. The prediction interval in the analysis is -0.54 to 1.71. This means that we cannot say with certainty that XAI always has a positive impact on human decision-making as the prediction interval is not exclusively larger than 0. 
Even though most studies report a performance improvement through XAI assistance, we also find studies that report a performance decline. In this context, one of the two studies conducted by \citet{alufaisan_does_2020} encountered the most negative effects with a human performance decline. Participants are asked to predict whether a defendant will recidivate in two years and receive AI predictions with decision rules aiming to support users' understanding of the AI's decisions. However, this reduction is not statistically significant due to a high level of dispersion. The most considerable improvement can be found in a study by \citet{bucinca_proxy_2020}. Here, participants have to decide based on an image of a meal whether the fat content of this meal on a food plate exceeds a certain threshold. 

It is important to highlight that the significant SMD does not imply that including explanations will improve performance over simply providing AI advice without any form of explainability, as we did not find a significant difference between AI-assisted and XAI-assisted performance in \Cref{sec:ai-vs-xai}. Instead, it can be interpreted as a positive effect of \textit{some} form of AI advice. 
\subsection{Tabular vs. Text Data} 

\begin{figure*}[h]
\centering
\includegraphics[width=0.95\linewidth]{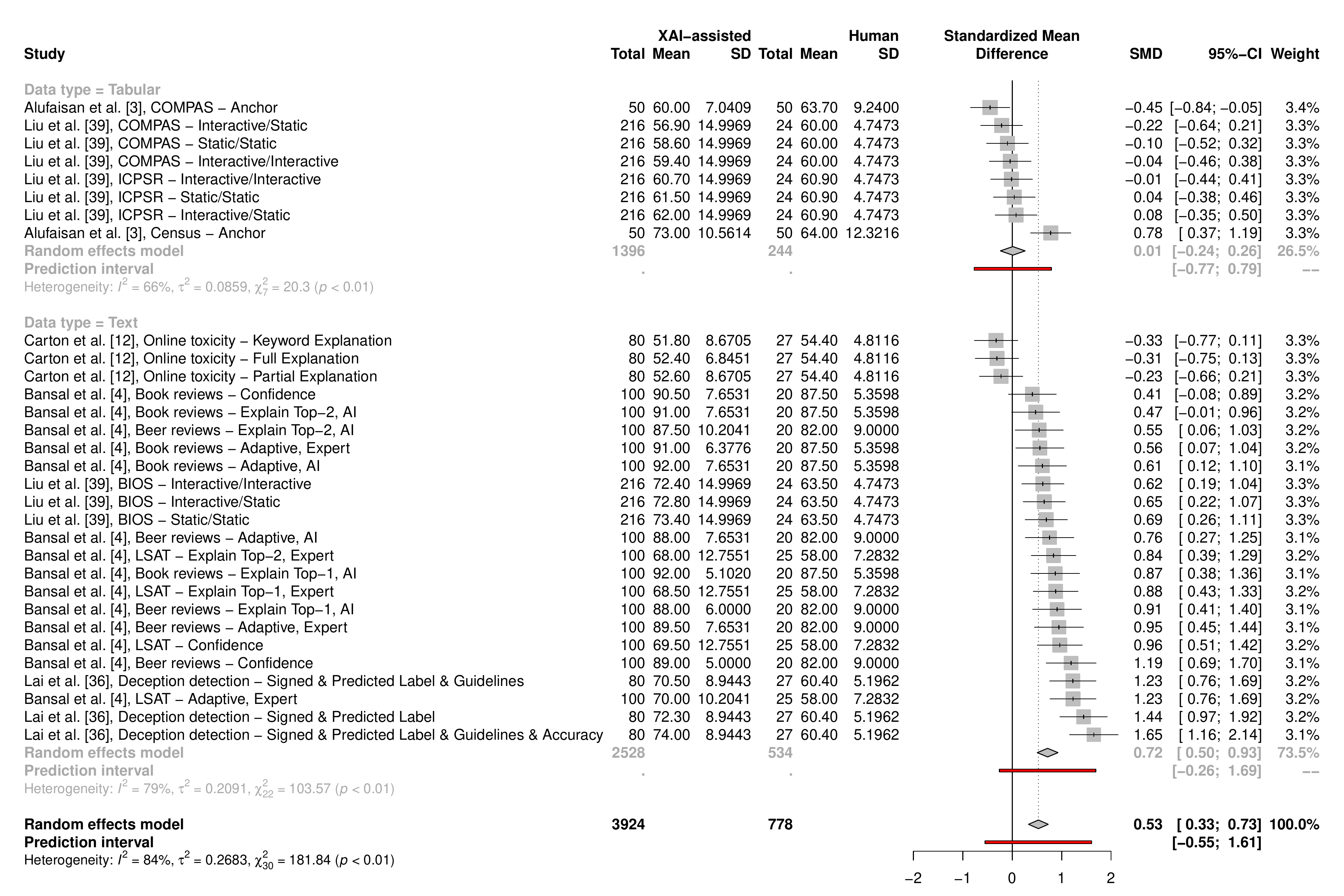}
\caption{Forest plot of the standardized mean difference between human and XAI-assisted performance considering the subgroups tabular and text data.}
\label{fig:subforest-datatype}
\end{figure*}

Additionally, we conduct a subgroup analysis based on three data types used in our sample---tabular, text, and image data. As only two articles report experiments using image data, the interpretation of this data type is not conclusive. Thus, we focus on tabular and text data types for the subgroup analysis resulting in a total sample size of 31 studies and 4,702 participants. \Cref{fig:subforest-datatype} displays the forest plot of the standardized mean difference between human and XAI-assisted performance with regard to both data types.

The SMD of the tabular subgroup is 0.01 (95\% CI [-0.24, 0.26]). As this range does include an effect size of 0 and the z-value is 0.08 with a corresponding p-value of 0.94, we cannot reject the null hypothesis. That means we cannot conclude that the SMD is significantly different from 0.
Regarding heterogeneity, we can reject the null hypothesis that the true effect size is identical in all studies (\(Q = 20.30, df = 7, p < 0.005\)). The \(I^2\) is 65.50\% with a 95\% CI ranging from 26.70\% to 83.80\% and \(\tau^2\) has a value of 0.09 (95\% CI [0.01, 0.48]). Finally, the prediction interval ranges from -0.77 to 0.79. That means we can expect future negative impacts of XAI assistance on human decision performance in certain situations. 

The text data subgroup has a higher SMD of 0.72 (95\% CI [0.50, 0.93]). This range does not include an effect size of 0. Additionally, the z-value is 6.65 with \(p < 0.0001\) denoting that the SMD is significantly different from 0. 
In terms of heterogeneity, we can reject the null hypothesis that the true effect size is identical in all studies (\(Q = 103.57, df = 22, p < 0.0001\)). In this context, \(\tau^2 \) is 0.21 (95\% CI [0.10, 0.47]) and \(I^2\) is 78.80\% (95\% CI [68.70\%; 85.60\%]). Thus, the heterogeneity of the text data subgroup can be considered higher than the tabular data subgroup. The prediction interval ranges from -0.26 to 1.69, which means we can also expect some negative XAI effects with text data.

Lastly, comparing both subgroups, we observe significant performance differences between tabular and text data suggesting that the data type in our sample influences the effect of XAI assistance on the performance (\(Q= 17.81, df= 1, p<0.0001\)). 

\subsection{Summary of the Articles}

Having analyzed the collected studies in the form of a meta-analysis, we pursue a discussion and summarization of the individual articles from a qualitative perspective. We focus on extracting further insights that can be derived from comparing human performance without any assistance and AI or XAI assistance. In particular, we are interested in the question of why XAI assistance did or did not improve AI assistance.

\textit{\citeauthor{alufaisan_does_2020}} \cite{alufaisan_does_2020} draw upon rule-based explanations generated by anchor LIME \cite{ribeiro2018anchors} for two real-world tabular datasets---an income prediction task using the Census dataset \cite{Dua2019} and a recidivism prediction task using the COMPAS dataset  \cite{propublica}.
The XAI algorithm provides rules which are denoted as anchors to explain a prediction on an instance level. A rule is considered an anchor if any changes in the features that are not included in the anchor do not impact the AI decision \cite{alufaisan_does_2020}.
The authors find that AI assistance improves human performance. However, they find no additional benefit of XAI assistance regarding decision-making performance. They hypothesize that one reason for the negligible effects of XAI could be information overload. When taking a closer look at the individual decision level, humans tend to follow AI predictions more often when the AI makes a correct prediction than when it makes an incorrect one. In this context, explanations did not alter this observation.

\textit{\citeauthor{bansal_does_2020}} \cite{bansal_does_2020} compare multiple forms of explanations of AI predictions on three tasks. Two of them are about sentiment analysis of book \cite{he2016ups}, and beer reviews \cite{mcauley2012learning}. The third task consists of a set of law school admission test questions requiring logical reasoning \cite{team2017lsat}. All three datasets consist of text data. 
As explanations, they use feature importance generated by LIME \cite{ribeiro_why_2016} and feature importance generated by human experts. Additionally, the authors evaluate the effect of providing only confidence ratings of the AI's prediction, which we consider as a form of explanation as well to ensure that AI assistance encompasses in all studies only the predictions of the AI. Therefore, we are restricted to the comparison of human performance with XAI assistance in the meta-analysis. Interestingly, the authors find no performance improvements of explanations over confidence ratings. However, they find that providing explanations tends to increase the chance that humans accept the AI's prediction regardless of its correctness---which shows contrary results to the observations of \citet{alufaisan_does_2020}. Moreover, participants receiving XAI assistance outperform both participants and AI alone.

\textit{\citeauthor{bucinca_proxy_2020}} \cite{bucinca_proxy_2020} analyze two different techniques for evaluating XAI systems. First, a proxy task, e.g., asking humans to predict the AI's decision. Second, an actual decision-making task, e.g., asking humans to make a decision with XAI assistance. We consider the actual decision-making task for the meta-analysis. The authors show participants images of different food plates and ask them to predict the fat content of the meal. The participants are requested to decide whether the fat content in percent is higher than a threshold. 
Users are provided with inductive (i.e., example-based) and deductive (i.e., rule-based) explanations. With XAI assistance, the users significantly outperform users with AI assistance, which yields better performance than users conducting the task without any support. Interestingly, example-based explanations enable users to identify erroneous AI predictions better. However, they prefer and trust the rule-based explanations more.

\textit{\citeauthor{carton_feature-based_2020}} \cite{carton_feature-based_2020} consider a social media comment toxicity prediction task for their experiment by sampling comments from the dataset provided by \citet{wulczyn2017ex}. 
They utilize feature importance by highlighting passages and words pointing towards a toxic comment. In detail, three different forms of explanations are analyzed---full, partial, and keyword explanations. The full explanations are generated by an attention model \cite{carton2018extractive} that produces a discrete attention mask over the input text for toxic content. Partial explanations refer to reducing the mask to the most toxic passage. Keyword explanations are derived from a bag-of-words logistic regression classifier that highlights only the most toxic single words instead of considering the context. 
Their study shows a marginal negative trend of AI assistance performance in comparison to users with no assistance at all. However, the effect does not vary significantly regardless of explanations being present or not. In this context, users tend to follow the AI prediction without an effect of the different explanations provided. A detailed analysis of false negative and false positive rates reveals that explanations tend to increase false negatives and reduce false positives. The authors hypothesize that this might indicate a reduced cognitive engagement with the social media comments by focusing on the highlighted text without considering the not highlighted passages.

\textit{\citeauthor{fugener2021will}} \cite{fugener2021will} use a dog breed image classification task. They evaluate the effect of AI assistance with and without additional confidence ratings and compare both treatments with the performance of users that do not receive any AI support. 
While the authors find that AI assistance significantly improves human performance, no additional effect of providing confidence ratings can be identified. In this context, both AI-assisted performance and the one with additional confidence ratings outperform humans and AI when conducting the task alone. Interestingly, a detailed analysis of the confidence rating treatment reveals that providing AI certainty decreases AI adherence. The authors hypothesize that a decrease in users' trust might explain this effect as it declines in this treatment as well.

\textit{\citeauthor{lai_why_2020}} \cite{lai_why_2020} consider a deception detection task which is about identifying fake hotel reviews \cite{ott2013negative,ott2011finding}.  
The authors analyze the effect of three types of XAI assistance provided together with the AI prediction. First, they provide users with signed feature importance, i.e., they highlight the most important words that indicate real and fake reviews. Word importance is derived from the absolute value of the coefficients using a linear SVM with a unigram bag-of-words. Second, they additionally provide the participants with supporting guidelines derived from related research and observations of the model, which are paraphrased by the authors. Third, they provide an additional AI accuracy performance statement on top. Before the actual decision-making task, participants had to undergo a task-specific training phase. In a prior experiment, the authors demonstrated that these so-called tutorials enhance participants' performance without any further assistance from the AI.  
In general, they find an improvement in XAI-assisted performance over human performance in all three treatments. However, no statistical significance between the different forms of XAI assistance can be detected. Moreover, the authors also emphasize that XAI-assisted performance remains inferior to the AI performing the task alone, even though a small proportion of participants can be identified that were able to outperform the AI. The authors do not consider an AI-assisted treatment in their experiment.

\textit{\citeauthor{liu_understanding_2021}} \cite{liu_understanding_2021} explore the effect of out-of-distribution data instances and interactive explanations on human-AI decision-making. For the meta-analysis, we focus on the in-distribution setting, including a comparison of static with interactive explanations, as out-of-distribution data might impede the comparability with the other studies. The authors draw upon three tasks in their article. In the first, participants predict whether arrested defendants will violate the terms of pretrial release using the ICPSR dataset \cite{ICPSR}. The second task is about predicting whether defendants will recidivate in two years using the COMPAS dataset \cite{propublica}. Both are tabular datasets. The third task requires participants to predict a person's profession given a textual biography using the BIOS dataset \cite{bios}. 
Regarding the AI model, a linear SVM classifier with unigram bag-of-words for BIOS and one-hot encoded features for ICPSR and COMPAS is employed. The static explanations consist of feature importance by coloring features that contribute to the AI prediction. Interactive explanations offer users the possibility to explore what-if scenarios. 
In general, the authors find that XAI assistance improves users' performance in predicting a person's profession compared to users without any assistance. However, no significant performance difference between static and interactive explanations can be found. For the two recidivism prediction datasets, they find no significant difference between the performance of XAI assistance and human alone. The authors hypothesize that the complexity of both recidivism tasks might have prevented noticeable performance improvements. Interestingly, even though no performance difference can be observed, users rate interactive explanations more useful in the recidivism prediction tasks. The authors do not consider sole AI-assisted decision-making in their article. 

\textit{\citeauthor{van_der_waa_evaluating_2021}} \cite{van_der_waa_evaluating_2021} evaluate the effect of example- and rule-based explanations in the context of a diabetes self-management task where participants are requested to select the appropriate dose of insulin. 
The authors compare AI assistance with both types of explanations but do not report sole human performance. In this context, the presence of either example- or rule-based explanations does not result in a significant performance difference compared to pure AI assistance. A closer analysis of explanations' ``persuasive power'', i.e., how often humans agree with the AI recommendation regardless of being correct or not, reveals that users without explanations follow the AI prediction significantly less than with example- or rule-based explanations.

\textit{\citeauthor{zhang_efect_2020}} \cite{zhang_efect_2020} compare the effect of additional confidence ratings with feature importance explanations using Shapley values \cite{lundberg2017unified} in the context of human-AI decision-making. They utilize the Census dataset \cite{Dua2019} for asking participants to predict whether a person's income would exceed \$50,000. 
In the experiment, the authors find no significant difference between additional confidence ratings and feature importance explanations in terms of task performance. Moreover, both treatments do not differ significantly from sole AI assistance. Even though displaying confidence scores does not affect task performance, it can be found that it improves overall trust and contributes to a calibration over different confidence levels. The authors explain this phenomenon with a high correlation between human and AI confidence, showing a large overlap of instances with low AI and human confidence. In contrast, explanations do not affect users' trust in their experiment. 

In summary, most studies, except for \citet{bucinca_proxy_2020}, found no significant differences when comparing XAI and AI assistance on an individual study level. \citet{alufaisan_does_2020} argue that this negligible effect of XAI might stem from information overload. \citet{carton_feature-based_2020} discuss that XAI, in the case of feature importance, might reduce the amount of data that humans process as they just focus on the features that are relevant for the AI. \citet{liu_understanding_2021} discuss whether task complexity could be a reason for no significant improvements of XAI assistance over AI assistance.
We also observe some contrasting results that require further investigation in future work. \citet{fugener2021will} find a general decrease in AI adherence in the context of confidence ratings. In contrast, \citet{van_der_waa_evaluating_2021} and \citet{bansal_does_2020} find that explanations just increase the general probability of accepting AI advice.

Based on the qualitative review, we derive two potential factors influencing the utility of XAI: First, XAI could improve decision-making performance by increasing the acceptance of AI advice. Note that the performance improvement will just happen if the AI, on average, performs significantly better than the human \cite{zhang_efect_2020}. Second, XAI assistance could influence appropriate trust and reliance, which means humans can discriminate between correct and incorrect AI advice \cite{bansal_does_2020}. The idea is that humans will be better able to distinguish between correct and incorrect advice if the AI conveys its reasoning. Our qualitative review showed that while there is evidence for increased acceptance of AI advice due to XAI \cite{bansal_does_2020,van_der_waa_evaluating_2021,zhang_efect_2020}, just one article reports some form of appropriate reliance \cite{bucinca_proxy_2020}. 

\section{Limitations}\label{sec:limitations}

As XAI is a relatively new field of research, at least in comparison to research domains where meta-analyses are more common, e.g., medical research \cite{eaden2001risk}, we encounter some major limitations that form around the current existing sample of XAI studies. 

First, the current existing sample of XAI studies contains just online studies. In these online studies, people are recruited via online platforms such as Mechanical Turk. They conduct a task not in a controlled lab environment inducing higher variability. Second, the studies use different XAI algorithms ranging from providing an additional confidence score to personalized explanations. We also considered a subgroup analysis for the XAI algorithm category but are limited due to the current sample size. Therefore, interpretable findings cannot be derived yet. Third, also task design differs between the studies. Some studies use more intuitive tasks for humans, such as sentiment analysis of reviews, while others consider more complicated ones, such as income prediction. Future studies should evaluate other task-related factors beyond data type. Fourth, in the data type subgroup, the tabular subgroup contains just two articles \cite{alufaisan_does_2020,liu_understanding_2021}. Even though it contains 8 studies, this poses a possible limitation.

Moreover, as many studies did not report dispersion metrics numerically, we needed to extract them from the plots. However, we conducted a multi-step approach. Two researchers extracted the values individually and afterward discussed the differences. Furthermore, we want to highlight that the meta-analysis is limited with regard to drawing causal conclusions. Our analysis should instead be considered as a synthesis of existing research.
The main limitation of the comparison of AI-assisted and XAI-assisted performance is the small sample size due to many studies not reporting the dispersion of their AI-assisted condition.

\section{Discussion and Future Work}\label{sec:discussion}
In this work, we conducted the first meta-analysis of XAI-assisted decision-making. Based on a structured data collection process, we collected 393 XAI-related articles. After applying our inclusion criteria, we identified a set of 9 articles encompassing 44 studies for the meta-analysis. 

In the current sample, we find no statistically significant difference between XAI- and AI-assisted performance. In this context, we observed that some studies reported positive XAI assistance performance effects, whereas others found no effects or even slightly negative effects. 

Additionally, we find a positive effect of XAI assistance on human task performance. Since we do not identify a difference between XAI and AI assistance, the results need to be interpreted carefully. Therefore, we cannot conclude that XAI will lead to an overall performance superior to AI assistance. However, we observe a positive tendency of AI to support humans in decision-making, either with explanations or without. A promising avenue for future research is to investigate the factors that determine consistent performance gains in human-AI decision-making. In this context, the work of \citet{lai_towards_2021} could be a foundation for the data collection.

Furthermore, our subgroup analysis indicates a stronger positive effect on task performance with text data than tabular data. If this effect can be confirmed in future studies, more work would be required regarding human-AI decision-making with tabular data. Reasons for this difference could be that text data is a more intuitive data type for humans. Additionally, analyzing the utility of explanations and AI predictions concerning image data requires more attention, as we found a total of only 2 articles that used image data. Future work should explicitly investigate performance differences induced by different data types. 

Moreover, due to the currently high heterogeneity of the studies, future analyses could consider not only a distinction by data type but also by task type, e.g., complexity, and users' task-specific knowledge. For example, in easy tasks, humans might be able to evaluate better whether AI advice is correct or not. 

Our qualitative review of the collected studies highlights that explanations can easily lead to increased acceptance of AI advice. In the scenario in which AI performs, on average, better than humans, its performance might bound the maximum joint performance of both. However, if the goal is that human-AI decision-making ideally results in superior team performance \cite{bansal_does_2020}, appropriate reliance on the AI's decision becomes indispensable. Therefore, future research should investigate design mechanisms that enable appropriate reliance. Additionally, from this observation emerges the need to discuss the ethical implications of pure acceptance increase in future work as it can be understood as a form of manipulating people to blindly accept AI advice.

\section{Conclusion}\label{sec:conclusion}
This article presents the results of a meta-analysis of the utility of XAI-assisted decision-making. We identify a total sample of 9 articles that report all necessary information as a prerequisite through a structured literature review. We analyze whether humans' decision-making can benefit from AI support with and without explanations and derive three major findings: First, we do not find a significant effect of state-of-the-art explainability techniques on AI-assisted performance. Second, we observe a significant positive effect of XAI assistance on human performance. Third, our analysis indicates that XAI assistance is more effective on text than tabular data.
We hope that our work will motivate scholars to pursue meta-analyses in future human-AI research to systematically assess previous studies to derive conclusions about the current body of research.

\bibliographystyle{ACM-Reference-Format}
\bibliography{references}

\end{document}